\documentstyle[11pt,aasms4,epsf]{article}

\def\folio{\ifnum\pageno<2\nopagenumbers\else\number\pageno\fi}
\newtoks\headline \headline={\hss\twelverm\folio\hss} 
\newtoks\footline \footline={{\hfil}} 
\font\mathbf=cmmib10 scaled 1000             

\def\ref{\par\noindent\hangindent=2pc \hangafter=1 }
\def\amin{\ifmmode^{\prime}\else$^{\prime}$\fi}
\def\asec{\ifmmode^{\prime\prime}\else$^{\prime\prime}$\fi}

\def\etal{{et al. }}
\def\cappage #1 #2 #3 {\vfill\eject\pageno=#1
\vglue 10 true in minus 10 true in \noindent{\bf Figure #2.} #3}
\def\ee #1 {\times 10^{#1}}
\def\ut #1 #2 { \, \hbox{#1}^{#2}}
\def\u #1 { \, \hbox{#1}}

\def\msol{\, \hbox{$\hbox{M}_\odot$}}

\let\grad=\nabla
\def\cross{{\bf \times}}
\def\curl #1 {\grad \cross #1}
\def\div #1 {\grad \cdot #1}

\def\msol   {\hbox{$M_\odot$}}                  
\def\etal   {{\it et al. }}                     

\begin{document}

\title{{\it HST} and {\it UKIRT} Observations of the Center of the Trifid 
Nebula: Evidence
for the Photoevaporation of a  Proplyd and a Protostellar
Condensation}

\author{F. Yusef-Zadeh}
\affil{Department of Physics and Astronomy, Northwestern University,
Evanston, Il. 60208 (zadeh@northwestern.edu)}

\author{J. Biretta}
\affil{STScI, 3700 San Martin Drive, Baltimore, MD 21218
(biretta@stsci.edu)}

\author{T. R. Geballe}
\affil{Gemini Observatory, 670 N. A'ohoku Pl., Hilo, HI 96720
(tgeballe@gemini.edu)}

\begin{abstract}

The Trifid nebula (M20) is a well-known prominent optical HII region 
trisected by bands of obscuring dust lanes and excited by an O7.5 star HD 
164492A.  Previous near-IR, mid-IR and radio continuum observations of the 
cluster of stars at the center of the Trifid nebula indicated 
circumstellar disks associated with hot stars with envelopes that are 
photoionized externally by the UV radiation from the hot central star, HD 
164492A. Using WFPC2 of the HST, we present evidence of a resolved proplyd 
in H$\alpha$ and [SII] line emission from a stellar source emitting cool 
dust emission. Using UKIRT, an infrared observation of the stellar source 
with a proplyd indicates a late F -- mid G spectral type. We also note a 
remarkable complex of filamentary  and sheet-like structures which appear 
to 
arise from the edge of a protostellar condensation. These observations are 
consistent with a picture in which the bright massive star HD 164492A is 
responsible for the photoevaporation of protoplanetary disks of other 
less-massive members of the cluster as well as the closest protostellar 
condensation facing the central cluster.  Using the evidence for a 
proplyd, we argue that the massive and intermediate mass members of the 
cluster, HD 164492C (B6 star) and HD 164492 (Herbig Be star) have disks 
associated with them.


\end{abstract}


\section{Introduction}

Circumstellar disks are an integral part of star formation. Their
existence has been established through a variety of millimeter, infrared 
and optical observations.  HST observations have played an important role
in transforming of our understanding of                      
circumstellar material distributed within 
or close to an HII region powered by a cluster of massive stars.
The so-called  ``proplyds'', or protoplanetary disks, and candidate 
proplyds 
have been found in
the Orion Nebula  (e.g., Laques \& Vidal 1979;  Garay et al.\ 1987; 
Churchwell et al.\
1987; O'Dell 
2001; Bally et al. 1998)  
in  the Lagoon nebula (M8)
(Stecklum et al. 1998), NGC 3603 (Brandner et al. 2000), Trifid nebula (M20) 
(Yusef-Zadeh, et al. 2000; Lefloch et al. 2002) and Carina nebula
(Smith, Bally and Morse 2003).  The Trifid nebula  is a
well-known HII region that is estimated to be $\sim 10^5$ years old 
and star formation has taken place at the edge of
the region within the last $10^4$ years (Cernicharo et al 1998; Lefloch
and Cernicharo 2000; Hester et al. 1999). It therefore presents us with an
opportunity to
examine proplyds at an  stage of their exposure to the ionizing
flux as well as  to investigate the type of stars that proplyds 
can be associated with. 

M20 is approximately 10$'$ in diameter with a distance estimated to be
at the distance of 1.68 
kpc  (e.g., Lynds et al. 1985; see also Kohoutek et al.
1999)  
and is centered on a small cluster of hot stars dominated by components A
through G of HD 164492.  The ionizing flux required to maintain the HII
region, $10^{48.8}$ s$^{-1}$, is supplied by the O7.5III star HD 164492A
(e.g., Walborn 1973; Yusef-Zadeh et al.  2000), which has M$_v$=--5.3 for
A$_v\approx$1.3 towards the central stars (Lynds and O'Neil 1985). Several
molecular condensations associated with protostellar sources lie within or at
the edge of the HII region suggesting that a new generation of massive star
formation has been induced by the expansion of the nebula (Cernicharo et al.\
1998; Lefloch \& Cernicharo 2000; Lefloch et al. 2002).

Radio and Infrared studies of the central region of M20 have suggested the 
presence of circumstellar disks in the vicinity of HD 164492A.  In one 
study, three compact radio continuum sources with thermal spectrum was 
detected within $0.''1$ of the cluster members identified by Gham et al. 
(1983)  as a A2 IA star (HD 164492B), a B6 V star (HD 164492C) and D 
(Yusef-Zadeh et al. 2000).  HD 164492D lies only 2$''$ west of star C and 
is classified as a Herbig Be-star LkH$\alpha$ 123 (Herbig 1957).  The D 
star is also reported to be the brightest point source at 12.5 and 17.9 
$\mu$m images (Yusef-Zadeh et al. 2000).  By analogy with M42, radio 
continuum emission from these 
sources was argued to be due to proplyds being photoevaporated by 
the intense 
UV 
field of HD 164492A (Yusef-Zadeh et al.\ 2000).  The case for this idea 
proceeded along the lines developed by Garay et al.\ (1987) and Churchwell 
et al.  (1987) for the 
proplyds first detected in radio continuum in M42. Another line of 
evidence for the existence of circumstellar disks came from near-IR L band 
observations (Yusef-Zadeh et al. 2000) showing several sources which were 
very dim or unseen shortward of K band image. Additional support for 
circumstellar disks surrounding the cluster members came from ISOCAM 
observations where several infrared sources were detected (Lefloch et al.  
2002).  These observations indicated two-temperature model of dust 
emission from mid-IR sources ranging between 150-200K and 500-1000K. The 
dust emission from HD 164492C and D implied the presence of warm neutral 
material
 around hot stars.
 Here, we present conclusive evidence for a proplyd from a  star
coincident with a mid-IR source IRS3 
based on the {\it Hubble
Space Telescope}
(HST) observations of the bright central region of M20. 


\section{Observations}
\subsection{HST}

WFPC2 observations (9104) of M20 were completed in June 2002 in order to make
a mosaic image of this nebula, to study proper motion of HH 399 jet in the
southeast region of the nebula and to search for proplyds in the central
region of the nebula.  Proper motion study of HH 399 was recently described by
Yusef-Zadeh, Biretta and Wardle (2004).  Deep narrow band observations with
overlapping exposures were made to make a mosaic image of the nebula.  A
detailed account of the large scale mosaic image will be given elsewhere.
Here, we concentrate on observations of the central region of the nebula
which contains a cluster of stars and a bright ionized rim known to be
embedded within the HII region.

To study the center of M20, a sequence of short and long exposures of the
central
cluster with two different orientations separated by 160$^0$ were carried out
in H$\alpha$ (F656N, 4$\times2$ s, 4$\times20$ s, 2$\times200$ s), [SII]
(F673N, 2$\times300$ s), [OIII] (F502N, 4$\times1.4$ s, 4$\times14$ s,
2$\times260$ s), F547M (2$\times40$ s), [NII] (F658, 4$\times1.4$ s,
4$\times14$ s) and [OI] (F631, 2$\times1$ s, 4$\times10$ s).  We used a range
of exposure times to insure non-saturated images for
the brighter stars, and used dithering to improve PSF sampling and
subtractions. The Planetary Camera (PC) chip was centered on the central
cluster for all the above observations. 

\subsection{UKIRT}

A 1.4-2.5~$\mu$m spectrum of the central star of HST1, as labeled in 
Figure 1,  was obtained at the
United Kingdom 3.8m Infrared telescope (UKIRT) on the night of 2004 Aug 29
(UT), using the facility spectrograph UIST (Ramsay Howat et al. 2000). The
0.24'' slit of UIST was oriented north-south and observations were made in
the standard stare/nod mode. The total exposure time was 4 minutes. Hipparcos
86814 (F6V) was used for rough flux calibration and removal of telluric
absorption lines.

\section{Results}

\subsection{Proplyd}

Figure 1 shows a grayscale PC image of H$\alpha$ line emission from
stars at the center of M20.  The five bright cluster members HD
164492A to E stars at the center of the nebula are labeled in Figure
1.  Two stars called HST1 and HST2 are located about 16.86$''$ and
9$''$ to the southwest of HD 164492A at $\alpha, \delta (J2000) = 18^h
02^m 23^s.53, -23^0 01' 51.87''$.  HST1 shows extended H$\alpha$ and
[SII] line emission centered at $\alpha, \delta (J2000) = 18^h 02^m
23^s.22, -23^0 01' 35.66''$.  A faint extended emission is also
detected from HST2 centered at $\alpha, \delta (J2000) = 18^h 02^m
23^s.07, -23^0 01' 45.45''$ but the reality of the extended component
needs to be confirmed.  Geometrical distortion has been removed before
the H$\alpha$ positions of these stars were measured, and the absolute
and relative positions of these stars are accurate to within 0.5$''$
and 0.01$''$, respectively.

Figures 2a,b show contour and grayscale representations of a close-up
H$\alpha$ view of HST1, respectively.  The morphology of the extended
emission is reminiscent of proplyds noted in M42 where a
crescent-shaped structure is pointing toward the central bright star
HD 164492A. The proplyd structure coincides within 0.22$^s$ and 
0.34$''$ in right ascension and declination   
of  a variable star centered at 
$\alpha, \delta (J2000) = 18^h 02^m
23^s, -23^0 01' 36''$
V3791 with its brightness varying between 12.9 and 15.1 magnitudes
(Kholopov 1998).  The integrated H$\alpha$ and [SII] fluxes from the
proplyd are estimated to be 4$\times 10^{-14}$ and 2.2$\times
10^{-15}$ erg s$^{-1}$ cm$^{-2}$, respectively. These estimates are made  
after the contribution
from the central star and the background emission have been
subtracted, but before any correction for extinction has been made.  In 
order to
investigate the reality of the extended emission and a dark ring
associated with HST1, a brighter field star was selected in the
vicinity of the central cluster.  The comparison of the field star with
HST1 showed that the number of counts from the crescent structure is
three times higher than the brightness of the diffracted ring detected
from the field star.  We also measured the number of counts from the
dark ring associated with HST1 and compared it with the background
emission.  We found the the brightness of the dark ring is about 2
counts pixel$^{-2}$ lower than expected from the diffraction of
starlight and the background.  The same crescent structure was noted
in two different exposures with different orientations as described
above.  These measurements suggest strongly that both the crescent
structure and the dark ring surrounding HST1 are real.  The outermost
and innermost radii of the crescent structure are
0.36$''$ and 0.22$''$ whereas the innermost radius of the dark ring is
$ < 0.14''$.

In order to identify the spectral type of the central star   HST1,
UKIRT observation of this star was carried out. 
Figure 3 shows the 2.0-2.5~$\mu$m portion of the spectrum. The spectrum 
has
been smoothed to a resolving power of $\sim$600. The Br~$\gamma$ line in the
calibration star was removed prior to ratioing; thus the Br~$\gamma$
absorption line in HST1 is real; it is the only clearly identified feature in
the spectrum. Its strength, the presence of weaker Brackett series lines in
the H portion of the spectrum, and the lack of CO overtone band absorption at
2.3~$\mu$m and longward suggest that the star's spectral type is late F - mid
G. Thus,  it is unlikely that the central star HST1 contributes a 
significant
fraction of the ionizing radiation for the proplyd.

In spite of the short PC1
exposures, HD 164492B, C and D stars were too bright to bring out  any
faint extended emission arising from these stars. However, we noted that 
both stars C and D show multiple components. The C star showed two 
bright peaks in H$\alpha$, [NI] and [OIII] images  separated by 0.1$''$ 
whereas the D star shows two bight peaks in H$\alpha$ separated by 0.1$''$,
one of which had an [OIII] counterpart. The high-resolution images suggest that 
HD 164492B, C and D stars are  binaries. 

Based on HST and UKIRT observations, we assume that HST1 has a 
protoplanetary disk that is being heated and
ionized by the FUV (6 eV $< h\nu < 13.6$ eV) and EUV ($h\nu > 13.6$
eV) radiation from HD 164492A. Assuming a projected distance of
2.8$\times 10^4$ AU (16.9$''$ at the distance of 1.68 kpc) between the
central star and V3791, the Lyman continuum flux J$_0$ is estimated to
be 2.8$\times10^{12}$ photons cm$^{-2}$ s$^{-1}$.  The morphology of
the proplyd shows that the radius of the ionization front R$_i$ is
subtending an angular distance of of 0.2$''$ or $\approx 340 AU$.  The
total Lyman continuum flux intercepted at the radius of the ionization
front is expected to be $\pi R_i^2 J_0 \approx 2.2 \times 10^{44}$ ph
s$^{-1}$ assuming that there is no visual extinction between the
ionizing source and the proplyd.  The observed H$\alpha$ flux is
estimated to be 1.5$\times10^{43}$ ph s$^{-1}$ assuming A$_v$=1.3
magnitudes along the line of sight to the Trifid.  A number of factors
may contribute to a discrepancy of a factor of 15 between the expected
flux intercepted by the protoplanetary disk and the extinction
corrected H$\alpha$ flux.  The discrepancy could be partly due to 
the differential extinction toward the central star and the proplyd. 
Alternatively, this discrepancy could be  due to 
a significant number of recombinations in the gas between the 
proplyd and the ionizing star. Considering that the proplyd lies in the 
bath of radiation and is closer to  the ionization 
front  of the HII region, it is plausible that the incident ionizing 
flux is different  than  that resulting from  the r$^{-2}$ geometrical 
dilution. 

Using the emission measure E (cm$^{-6}$ pc) =4.9 $\times 10^{17} \rm
I_{\alpha} \approx \rm n_e^2$ L (see Bally and Reipurth 2001), the
electron density n$_e (\rm cm^{-3}$) at the ionization front can be
estimated from the observed H$\alpha$ intensity I$_\alpha\approx
2\times 10^{-13}$ ergs s$^{-1} \rm cm^{-2} arcs^{-2}$. 
 Assuming that L is the radius of the cusp $\approx 2.5 \times 10^{-3}$ pc 
or
504 AU (0.3'' at the distance of 1.68 kpc), n$_e \approx 1.1\times 10^4
$ cm$^{-3}$ after accounting for the extinction between us and M20.
Given this emission measure, the estimated radio continuum flux
density is estimated to be $\approx$ 0.1 mJy at 8GHz assuming the
temperature of 8000 $^0$K and the conversion factor used by Brandner
et al.  (2000).  The total mass of ionized gas is estimated to be
M$_{HII}\approx 4/3\, \pi\ \mu\ n_e\ R_i^3 \approx 2.1\times 10^{-5}
\msol $ where the mean molecular weight per electron $\mu$ is assumed
to be 1.4 $m_H$.  Estimating the expansion time to be $R_i / (10
\textrm{km/s}) \approx 160$ years, the mass-loss rate is estimated to
be $ 1.3 \times 10^{-7} \msol yr^{-1}$.

\subsection{Photoevaporative Ionized Rim}

Evidence for the support of photoevaporative process being at work
on a large scale comes from the striking distribution of the ionized gas 
$\approx25''$ away from HD 164492A. 
Figure 4 shows an HST  mosaic image of the central region of  M20. 
The mosaic image  is constructed by combining H$\alpha$, [SII] and [OIII]
line images. Two 
 bright ionized rims to the north and southeast delineate  the 
surface of  molecular condensations facing the central cluster. 
The so-called  molecular consdensation TC1  (the nomenclature used by 
Cernicharo et al. 1998) to the north is  the closest protostellar condensation
to the central cluster in M20. 
 Figure
5a,b
show a close-up view of the head of TC1  with  a 
cometary morphology.
A  remarkable
complex of elongated  structures parallel and perpendicular to 
 the ionized rim is best  noted  in H$\alpha$ and [OIII] line images. 
These features have no [SII] counterparts. The elongated structures extend up to 
12$''$ or $\sim$0.1 pc,  some of which appears to show  a helical structure in 
projection and some 
sheet-like appearance. 
The lack of [SII] line emission from the filaments away from the ionized
rim  may be the result of   an  ionization stratification noted in the
photoionized
photoevaporative
flow as seen in the HST image of  M16 (e.g., Hester et al. 1996).   
Unlike the photoevaporative flows streaming normal to the surface of 
the dark cloud toward the ionizing star, the features noted in 
Figures 4 and 5  make 
 a large angle with respect to  normal to the surface of the 
molecular condensation. 


\section{Discussion}

The evidence for the existence of protoplanetary disks at the center of 
M20 being illuminated by a bright O7.5 star HD 164492A comes from a number 
of direct and indirect measurements.  The most conclusive evidence comes 
from the optical morphology of H$\alpha$ and [SII] line emission arising 
from a crescent-shape structure with its cusp pointing toward HD 164492A 
star. The proplyd coincides with  a late F -- mid G spectral type
and an infrared source IRS 3 which shows  evidence 
of warm dust emission from the protoplanetary disk.  
The physical size or the cusp radius  of the proplyd is 
$\approx8\times10^{15}$ cm which is larger than the typical 
size of the proplyds  
found in the Orion 
nebula . Radio continuum, near-IR and mid-IR measurements suggest dense 
material in the form of protoplanetary disks being photoevaporated by HD 
164492A. These stars include a massive star HD 164492C or IRS 1 and an 
intermediate mass star HD 164492D or IRS 2. The evidence of proplyd from 
IRS 3 provides indirect support for the hypotheses that the disks of these 
massive stars IRS 1 and IRS 2 are also photoevaporated by HD 164492A but 
because of their high stellar brightness, their corresponding proplyds were not 
detected.

One of the key questions is whether or
not proplyds  survive sufficiently long for planet formation to occur
(Churchwell et al.  1987, Henney \& O'Dell 1999).  To examine the
properties of the proplyd in M20, we estimate the lifetime of the neutral 
disk using the column density N(HI) $\approx 10^{21} \rm cm^{-2}$ 
estimated  from mid-IR observations (Lefloch et al. 2002). 
These authors derive a typical column density of hydrogen for the 
cool component of the dust emission from IRS 1, 2 and 3, all of which show  
similar spectral characteristics. If we assume that this 
cool column of gas comes from a disk of radius R$_H\approx0.14''$ or 235 AU, 
the total mass of the disk M$_{H} \approx N_{H}\ \pi\ R_H^2\ 
m_{H2} \approx 8.8\times10^{-5} \msol $. 
The mass of cold neutral  material is estimated to be similar to
the mass of ionized gas,  thus the emission is argued to be
originating from the photon-dominated region PDR of a reservoir of dense
circumstellar material exposed to the ionizing radiation of HD 164492A
(Yusef-Zadeh \etal 2000; Lefloch \etal 2001).
The  lifetime of the PDR  in the proplyd associated with 
IRS 3 is estimated to be roughly 500  years. However, this timescale is 
clearly underestimated since the column density of molecular gas is unknown.

Additional evidence for the support of photoevaporative process being at 
work on a large scale comes from the distribution of a bright rim seen 
$\approx36''$ away from HD 164492A. The bright ionized rim delineates the 
surface of a molecular condensation (TC1) facing the central cluster.  
Figures 4, 5 show a remarkable extended structure projected against the 
surface of the ionized rim as best noted in H$\alpha$ and [OIII] lines. 
This shock-like filamentary structures with size scale of about 0.1 pc are 
detected only at the edge of the ionization front where the incident 
radiation from HD 164492A in non-normal to the surface of the molecular 
condensation. We speculate that the origin of such a structure could be 
due to a number of proplyd outflows distributed in the vicinity of TC1. 
The nature of several stars projected against the filamentary structure 
needs further investigation to determine their relationship to  the 
filamentary structure. In this picture the proplyd outflows could be 
shocked by their interaction with each other and/or with the stellar wind 
from HD 164492A. Similar processes are known to take place in the Orion 
nebula (e.g., Garcia-Arrendono, Henny \& Arthur 2001; Henney 2002; Graham 
et al. 2002). Other effects that could contribute to the production of 
filamentary [OIII] structure at such curious location with respect to the 
central HD 164492A star is the onset of instability of the ionization 
front. Recent numerical work 
suggests that the ionization front can become unstable when the source of 
radiation makes an inclined angle with respect to the surface of TC1 
(e.g., Williams 2002). Obviously, more detailed work is necessary to 
understnad the nature of these diffuse ionzied features.


Acknowledgments:  We wish to thank R. O'Dell and Mark Wardle who were
co-investigators on the HST proposal under which these observations were 
carried out.  
We are also grateful to the referee, William Henney whose comments
improved  the manuscript.  In particular, his suggestions to account for 
the origin of the large-scale diffuse features in the vicinity of the 
protostellar 
condensation were extremely helpful. 
TRG thanks the staff of UKIRT; his research is supported by the Gemini
Observatory, which is operated by the Association of Universities for
Research in Astronomy on behalf of the international Gemini partnership of
Argentina, Australia, Brazil, Canada, Chile, the United Kingdom, and the
United States of America.

\begin{figure}
\caption{PC1 image of H$\alpha$ emission from the central cluster of 
stars in M20 based on 2$\times200$ sec exposures. The reference pixel
position is at 
$\alpha, \delta (J2000) =  18^h 02^m 23^s.2, -23^0 01' 38''$. The
labeled stars A to E 
correspond to cluster members HD 174492A -- E.}

\end{figure}
   

\begin{figure}
\caption{(a) Grayscale representation of the H$\alpha$ distribution of 
HST1 using PC1 [left panel]. (b) Contours
of H$\alpha$ emission from the protostellar disk with levels at (18, 20, 22,
24, 26, 28 and 30) $\times 
 2.051\times 10^{-17} \rm erg \rm cm^{-2} \rm s^{-1}$ }
\end{figure}

\begin{figure}
\caption{The 2.0-2.5~$\mu$m spectrum of the central star of HST1, shown at
R$\sim$600. The location of the Br~$\gamma$ line is indicated.}
\end{figure}

\begin{figure}
\caption{H$\alpha$, [SII] and [OIII] images of the inner 3$'$ of
M20.  Two bright 
ionized rims associated with two molecular condensations are noted 
to the north and southeast of the stellar  cluster at the center of the nebula.}
\end{figure}

\begin{figure}
\caption{(a) H$\alpha$  (a)  and [OIII] (b) images of the northern region
of  the bright ionized rim of the Trifid nebula where filamentary structures 
are detected.}
\end{figure}


\end{document}